# Portable Intrinsic Gradiometer for Ultra-Sensitive Detection of Magnetic Gradient in Unshielded Environment

Rui Zhang, [1,a)] Rahul Mhaskar, [2)] Ken Smith, [1)] and Mark Prouty [1)]

[1]*Geometrics, Inc., San Jose, CA, 95131, USA*

[2] *formerly Geometrics, Inc., San Jose, CA, 95131, USA, now Keysight Technologies, Santa Clara, 95051, USA*

We demonstrate a portable all-optical intrinsic scalar magnetic gradiometer composed of miniaturized cesium vapor cells and vertical-cavity surface-emitting lasers (VCSELs). Two cells, with an inner dimension of 5 mm x 5 mm x 5 mm and separated by a baseline of 5 cm, are driven by one VCSEL and the resulting Larmor precessions are probed by a second VCSEL through optical rotation. The off-resonant linearly polarized probe light interrogates two cells at the same time and the output of the intrinsic gradiometer is proportional to the magnetic field gradient measured over the given baseline. This intrinsic gradiometer scheme has the advantage of avoiding added noise from combining two scalar magnetometers. We achieve better than 18 fT/cm/√Hz sensitivity in the gradient measurement. Ultra-sensitive short-baseline magnetic gradiometers can potentially play an important role in many practical applications, such as nondestructive evaluation and unexploded ordnance (UXO) detection. Another application of the gradiometer is for magnetocardiography (MCG) in an unshielded environment. Real-time MCG signals can be extracted from the raw gradiometer readings. The demonstrated gradiometer greatly simplifies the MCG setup and may lead to ubiquitous MCG measurement in the future.

Over the last two decades, many breakthroughs have been achieved in atomic magnetometer research. For example, the discovery of the spin-exchange relaxation-free (SERF) phenomenon at high atomic density and low magnetic field leads to a great improvement in the magnetometer noise performance [1]. Sensitivities comparable with [2] or even outperforming [3] those of superconducting quantum interference devices (SQUIDs) have been reported with SERF magnetometers. Another example is the successful fabrication of atomic magnetometers using the technique of Micro-Electro-Mechanical Systems (MEMS) [4, 5, 6, 7, 8]. MEMS techniques enable chip-scale devices, significantly reducing size and power-consumption of atomic magnetometers. Chip-scale magnetometers can have sizes approaching 10 mm$^3$ and dissipate less than 200 mW. Despite all these advances, applications of atomic magnetometers are still limited. Highly sensitive SERF magnetometers require a magnetically shielded environment while chip-scale total-field magnetometers have subpar noise performances [5], although a scalar magnetometer with a sensitivity of 100 fT/√Hz has been demonstrated using a MEMS-based cesium vapor cell [9]. With bigger cells, scalar magnetometers can reach sensitivities of sub-10 fT/√Hz [10] or even sub-fT/√Hz [11]. In practical applications in an unshielded environment, the output noise of scalar magnetometers is often dominated by the background field fluctuation, instead of their sensitivities. To overcome this problem, a common solution is to set up a gradiometer system using two or more magnetometers [12, 13, 14, 15]. By taking the reading difference between adjacent magnetometers, this conventional gradiometer configuration suppresses the common field fluctuations at the cost of worsening the sensitivity by at least a factor of √2, compared with that of an individual magnetometer. Here, we describe an all-optical intrinsic scalar gradiometer scheme that reads out the magnetic gradient optically and thus avoids the reduced sensitivity issue with the conventional gradiometer. In addition, we built a portable device based on the gradiometer scheme, using miniaturized cesium vapor cells and vertical-cavity surface-emitting lasers (VCSELs), two components essential for compact and low-power devices. We demonstrate a gradiometer output noise density of less than 90 fT/√Hz, which is equivalent to 18 fT/cm/√Hz sensitivity in the gradient measurement for a baseline of 5 cm. A similar sensitivity is expected for a proposed atomic magnetometer system [16] and a slightly better sensitivity was achieved with a portable gradiometer using two scalar magnetometers recently [17].

a) Email: rzhang@geometrics.com

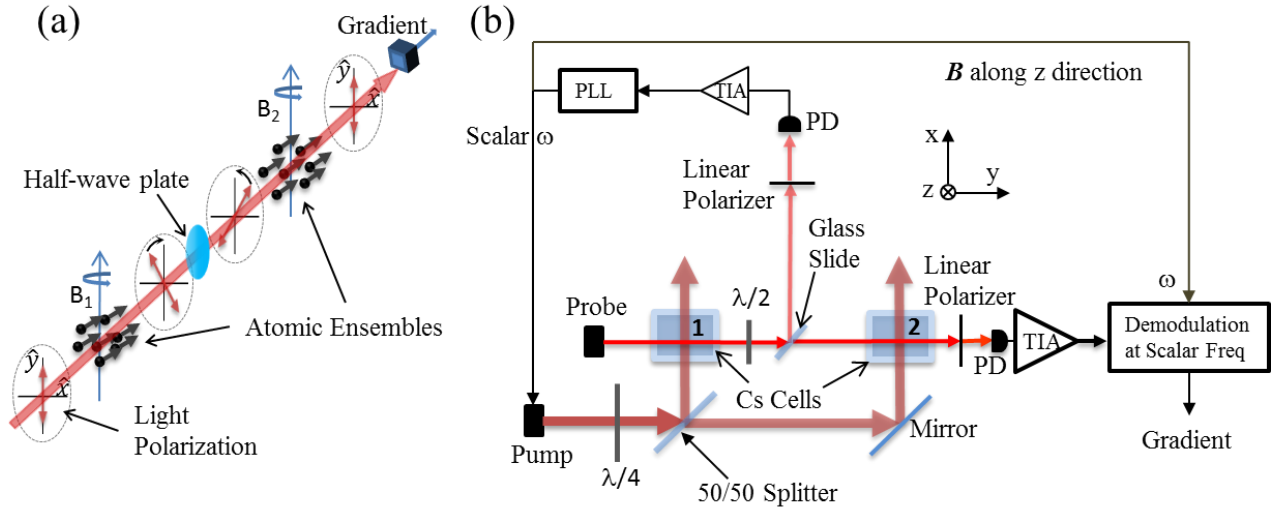

Figure 1 (a) Basic principle of the intrinsic gradiometer. (b) Schematics of implementing the intrinsic gradiometer.

The basic principle of the intrinsic gradiometer is illustrated in Figure 1 (a). A single linearly-polarized off-resonant probe beam interrogates two atomic ensembles separated by a baseline distance. The atoms at both locations are spin-polarized the same way and undergo Larmor precession. After interacting with the first atomic ensemble, the probe beam acquires the precession information in its polarization through optical rotation. A half-wave plate then reverses the phase of the oscillation in the probe polarization by 180 degrees. If the magnetic fields at the two locations are the same, the second ensemble of atoms rotates the probe polarization exactly the same way as the first and opposite to the existing polarization rotation of the incoming probe beam due to the half-wave plate. The optical rotation effect is cancelled. However, if there is a magnetic field gradient, the difference in the precession frequencies will result in an out-of-phase component in the rotation of the probe polarization with respect to the Larmor frequency of the first atomic ensemble. Therefore, the demodulated probe polarization signal at the Larmor frequency of the first atomic ensemble is a direct measurement of the magnetic field gradient. A similar optical read-out scheme was explored in the SERF regime [18].

The intrinsic gradiometer can be realized schematically shown in Figure 1 (b). Light from a modulated pump VCSEL is circularly polarized and then equally split into two beams, exciting cesium atoms at two vapor cells. A linearly polarized off-resonant beam from the probe VCSEL interacts with the two atomic ensembles at the same time. After a half-wave plate, about 10% light is split from the main probe beam by a piece of glass slide and is detected by a photodiode after an analyzing linear polarizer with its transmission axis at 45 degrees with respect to the input light polarization direction. The photodiode signal from the transimpedance amplifier (TIA) is demodulated by the phase detector within the phase-lock-loop (PLL) at the pump modulation frequency. The demodulated phase output is processed by the control loop within the PLL. The output of this control loop drives the pump laser through current modulation to match the Larmor frequency at the first cell. After the second cell, the main probe beam is analyzed by another linear polarizer with its transmission axis at 60 degrees with respect to the input light polarization direction and detected by a photodiode, whose current goes through another TIA. The TIA output is sent to a demodulator with its reference frequency determined by the PLL. The quadrature component of the demodulated probe signal is converted to the gradiometer reading with a constant determined by the slope of the quadrature component at the magnetic resonance.

We built a portable gradiometer by housing all optics and VCSELs inside a 3D printed structure with a dimension of 5 cm x 10 cm x 15 cm. Two cesium vapor cells with 60 Torr $N_2$ buffer gas and an inner dimension of 5 mm x 5mm x 5mm are separated by 5 cm. Two VCSEL packages, photodiodes and their transimpedance amplifier circuits are integrated on a printed circuit board (PCB), which is mounted on one side of the sensor head. The cells are 14 cm away from the PCB to avoid the magnetic contamination from the electronics. With customized non-magnetic electrical components, it is straight forward to reduce the sensor size to 3 cm x 6 cm x 3 cm. In operation, the pump VCSEL laser is resonant on the $|F=3\rangle \rightarrow |F'=4\rangle$ transition of the cesium D1 line and its current is modulated by the PLL output. The pump beam has a total power of 350 μW. The probe VCSEL is 15 GHz red detuned with respect to the $|F=4\rangle \rightarrow |F'=3\rangle$ transition of the cesium D1 line with 150 μW in power. Both lasers have a Gaussian spatial profile with FWHM of about 1.8 mm. The vapor cells are heated to 68°C.

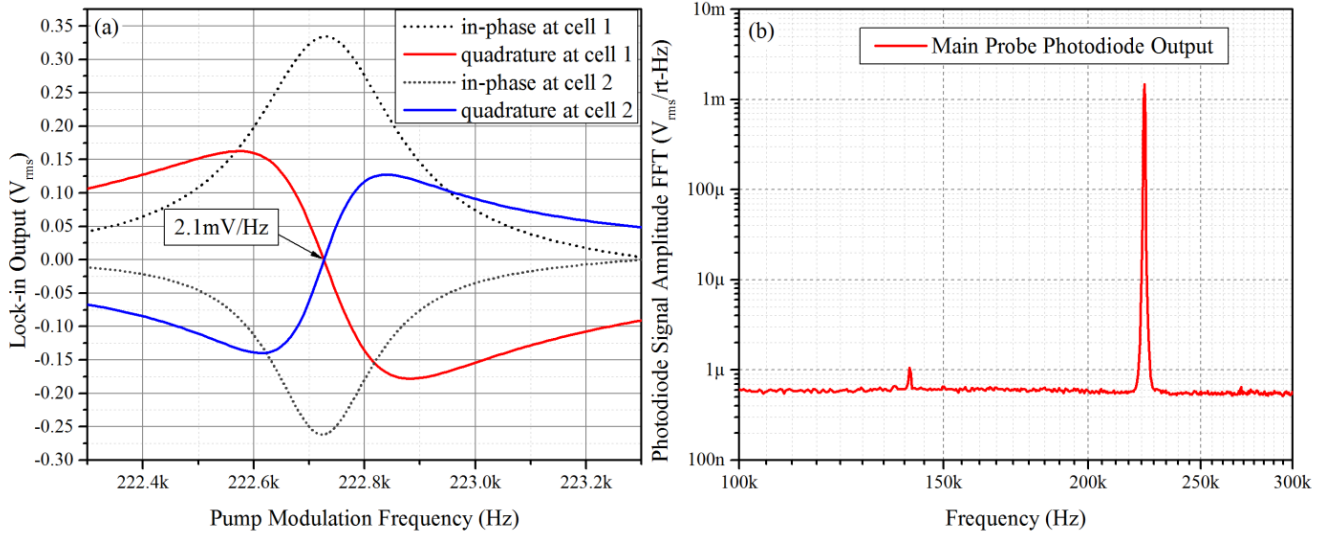

Figure 2 (a) Main probe lock-in signal as a function of pump modulation frequency. The black dot (red and blue solid) curves represent the in-phase (quadrature) components at two cells. (b) Amplitude spectral density of the main probe photodiode signal after the transimpedance amplifier.

We first test the gradiometer inside a shield can with the background magnetic field set at 63,600 nT. Magnetic resonance at an individual cell can be measured by demodulating the main probe signal at the pump modulation frequency while blocking the pump beam exciting the other cell. Both the in-phase and the quadrature components at two cells are shown in Figure 2 (a). As seen, the precession signals at the two cells are exactly opposite in phase due to the half-wave plate. The cell temperatures are adjusted to optimize and match the slopes of the two quadrature components at the magnetic resonance frequency. Relative cell temperature fluctuation affects the slope matching and can increase gradiometer output noise, especially at low frequencies. With less than 10 mK temperature fluctuation in our cells, we do not observe obvious cell temperature related noises. The optimized slope is $k = 2.1$ mV/Hz. We also measure the main probe photodiode noise spectral density, $d \approx 550$ nV/√Hz, as shown in Figure 2 (b). The individual magnetometer noise density can be estimated as $d/(k*\gamma) \approx 75$ fT/√Hz, where $\gamma \approx 3.5$ Hz/nT is the gyromagnetic ratio of cesium atoms. The gradiometer noise is the same as that of the individual magnetometer as long as the gradient is much smaller than the width of the magnetic resonance. Similarly, the scalar magnetometer using the split probe beam is estimated to have a sensitivity of 300 fT/√Hz.

The sensitivity of atomic magnetometers is often limited by the photon shot noise, which is given by $\sqrt{(2h\nu p)}$, where $h\nu$ is the single photon energy and $p$ is the total light power. After the linear polarizer, the main probe power is reduced to 26 μW. With $\lambda = 895$ nm, photodiode efficiency of 0.65 A/W and 200 kΩ transimpedance amplifier gain, the spectral density of the main probe photodiode signal due to the photon shot noise is 440 nV/√Hz, which is close to the measured value at 550 nV/√Hz. This indicates that the sensor operates close to the photon-shot-noise limit.

The scalar magnetometer based on the split probe beam operates in the closed-loop mode and measures the magnetic field at the first cell. The main probe signal, carrying the gradient information, is demodulated by the scalar magnetometer output frequency. The phase of the demodulator is set to be the same as that of the PLL for the split probe signal. The quadrature output of the demodulator is converted directly to the gradient by multiplying a constant given by $1/(k*\gamma) \approx 0.1361$ nT/mV. The gradiometer output is essentially the combined signal of the two quadrature components, as shown in Figure 2 (a), at a certain pump modulation frequency determined by the scalar magnetometer. The gradiometer sensitivity is not affected by the fundamental noise of the scalar magnetometer, which is equivalent to a small fluctuation of the pump modulation frequency around the actual magnetic resonance frequency. As seen in Figure 2 (a), as long as the slopes match, two quadrature components still cancel each other even when the pump modulation frequency fluctuates. The better the slopes match, the less affected the gradiometer sensitivity is by the fundamental noise of the scalar magnetometer.

The noise densities, averaged for 6 minutes, of both the scalar magnetometer (red) and the intrinsic gradiometer (green) are shown in Figure 3. As seen, the scalar output noise density is 300 fT/√Hz as expected, while the gradiometer noise density is below 90 fT/√Hz, which is close to the estimated noise density of 75 fT/√Hz. The additional gradient noise observed may be due to the magnetic shield [19], since the vapor cells are only 5 cm away from the shielding metal considering the size of the shield can. The advantage of the gradiometer scheme is clear even at the noise level of 90 fT/√Hz, since the difference of two 75 fT/√Hz scalar magnetometers would exhibit noise of at least 106 fT/√Hz. Another advantage is that the noise of the scalar reference channel cancels as common mode in the gradiometer measurement as long as the two slopes shown in Figure 2 (a) match. This considerably simplifies the design of the sensor driver since the signal-to noise ratio (SNR) requirements of both the pump laser current, as well as the pump current modulation can now be relaxed.

The accuracy of the gradient conversion constant is verified by a 20 Hz testing signal. The signal is generated by a 3 mm diameter coil attached to the first cell. Due to the size of the coil and the distance between the coil and the second cell, the signal is almost negligible at the second cell. Therefore, the 20 Hz signal measured by the gradiometer should have the same amplitude as that measured by the scalar magnetometer if the conversion constant is accurate, which is confirmed as shown in Figure 3. By sending fixed amplitude signals at different frequencies through the same 3 mm coil, the 3-dB bandwidth of the scalar magnetometer and the gradiometer is measured to be 300 Hz and 150 Hz, respectively. The bandwidth data is plotted in the inset of Figure 3. The higher bandwidth of the scalar magnetometer is due to the control loop parameters of the PLL, which extends the frequency response while increasing high-frequency noise.

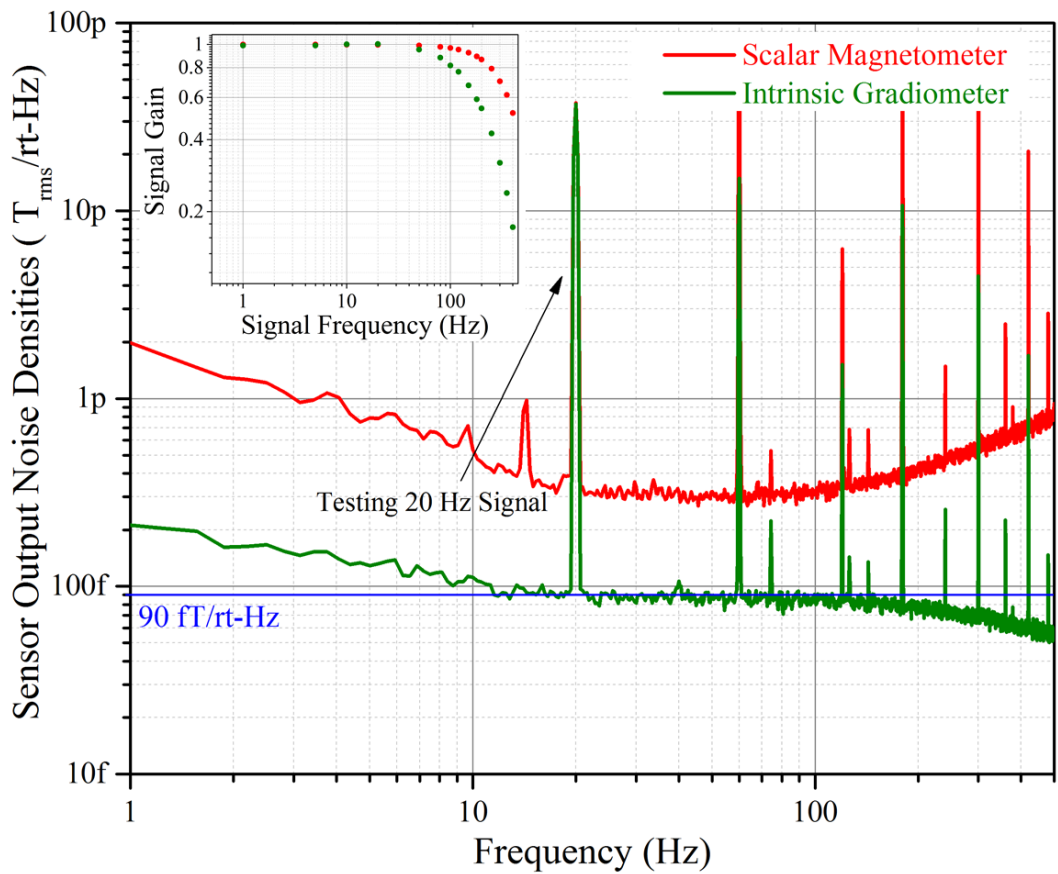

Figure 3 Noise densities of the scalar magnetometer (red) and the intrinsic gradiometer (green) inside a shield can. The inset shows their bandwidths.

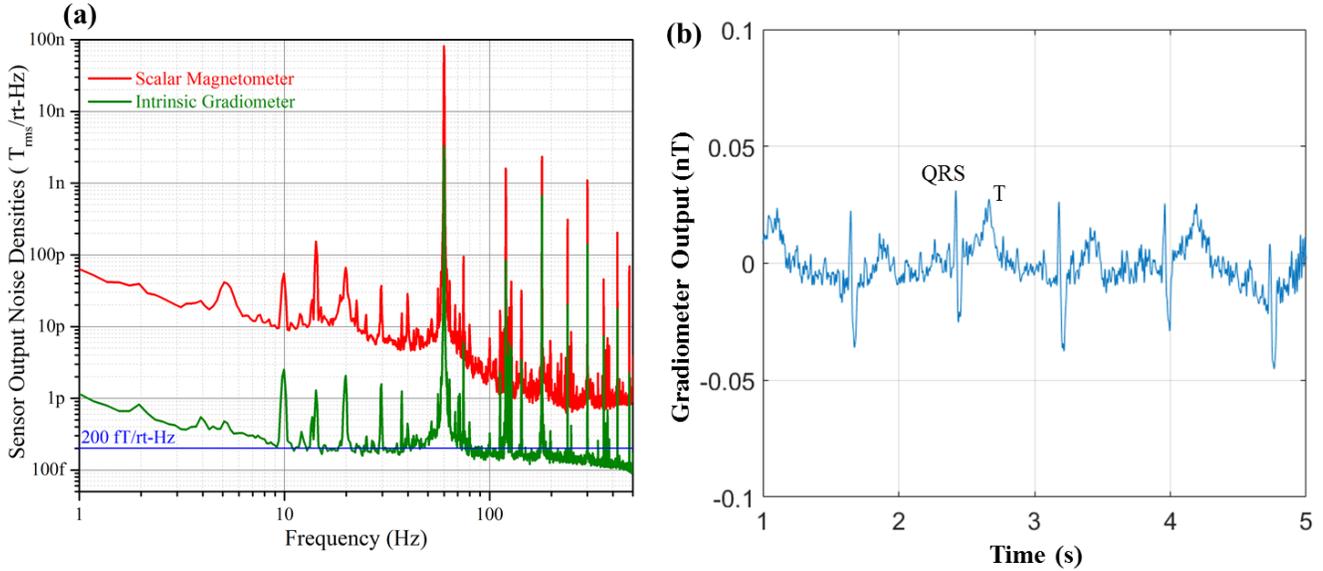

Figure 4 (a) Noise densities of the scalar magnetometer (red) and the intrinsic gradiometer (green) in an unshielded environment. (b) Real-time MCG data with 60 Hz notch filter and 70 Hz low-pass filter.

Accounting for the baseline, the gradiometer sensitivity is better than 18 fT/cm/√Hz. With the bandwidth of over 100 Hz, the gradiometer can be used for many applications such as nondestructive evaluation [20], unexploded ordnance (UXO) detection [21] and magnetocardiography (MCG) in an unshielded environment [22, 23, 24, 25]. The short 5cm baseline of the gradiometer compared to relevant geophysical length scales of meters allows dark-field UXO detection since the sensitivity of the gradiometer enables the measurement of local magnetic gradient signal due to the UXO while the dynamic common-mode environmental noise is cancelled. The analysis of SNR in detecting various types of UXO under different geophysical conditions is outside the scope of this paper and warrants further investigation.

Here we demonstrate the MCG application in an office environment. The noise density in the unshielded environment is first measured and plotted in Figure 4 (a). The gradiometer noise is worse compared with that in the magnetic shield due to the real gradient fluctuation inside a commercial building. From the noise density plot, it is also obvious that we need to remove the 60 Hz power line noise and its harmonics, which is done by a 60 Hz notch filter and a 70 Hz low-pass filter. The MCG data is recorded when a person is standing by the gradiometer with his chest 2 cm away from the second cell. After the two filters, the gradiometer output is shown in Figure 4 (b). As seen, even in a very magnetically noisy office environment, real-time MCG signals are clearly visible.

In conclusion, we demonstrate a portable all-optical intrinsic gradiometer that can be operated in unshielded environments. By using miniaturized vapor cells and VCSEL lasers, we show the feasibility of an ultra-sensitive magnetic gradiometer in a highly compact physics package. Combined with the all-optical operation to avoid the interference between nearby devices, such a sensor can be built into arrays for fast data collection and further magnetic noise suppression. For applications involving mobile platforms, such as UXO detection, more research efforts are necessary to study heading error effects. One of the main issues of the current gradiometer scheme is the dependence of the gradient conversion constant on the relative orientation between the sensor and the background magnetic field, causing a large heading error. In principle this heading error can be minimized by introducing an automatic gain based the amplitude of the input precession signal. As long as the quality factor of the magnetic resonance remains the same, the conversion constant will not change. On the other hand, we believe that this work eliminates major challenges of applying atomic magnetometers for stationary applications, such as MCG in unshielded environments. Development of commercially available intrinsic gradiometers can definitely advance the MCG research and may lead to ubiquitous MCG measurement in the future.

The data that support the findings of this study are available from the corresponding author upon reasonable request.

The authors would like to thank Mehdi Malek and Ken Glasser for their help in designing the 3D gradiometer housing and Lynn Edwards for his work on setting up testing fixtures. This work is supported by Strategic Environmental Research and Development Program (SERDP) under project number MR19-1379.